\documentclass[aps,prl,twocolumn,groupedaddress]{revtex4-2}

\usepackage{amsmath}
\usepackage{url} 
\usepackage{graphicx}
\usepackage{xcolor}

\begin{document}

\preprint{https://doi.org/10.1002/lpor.202501775}

\title{All-fiber microsensor of polarization at~single-photon level aided by deep learning}

\author{{Martin} {Bielak}${}^\dagger$}
\author{{Dominik} {Va{\v{s}}inka}${}^\dagger$}
\author{{Miroslav} {Je{\v{z}}ek}*}

\affiliation{{Department of Optics, Faculty of Science}, {Palack\'{y} University},\\ 
	{{17.~listopadu 1192/12}, {77900 Olomouc}, {Czechia}}\\
	${}^\dagger$ These authors contributed equally to this work\\
	* Email Address: {jezek@optics.upol.cz}\\
	\url{https://doi.org/10.1002/lpor.202501775}}

\date{\today}

\begin{abstract}
	
	Polarization of light carries vital information in numerous scientific disciplines, including biomedical imaging, optical diagnostics, and environmental sensing. However, accurate polarization measurements in constrained spaces, under low-light conditions, and at high speeds remain a severe challenge. We present a compact, all-fiber polarization sensor capable of single-shot, real-time operation with single-photon sensitivity and long-term stability. The sensor leverages intermodal interference within a short segment of a few-mode fiber, coupled to an array of single-photon detectors. The recorded detections are processed by a neural network, enabling precise reconstruction of complete polarization information for fully and partially polarized states. This robust architecture allows for thousands of polarization state measurements per second while achieving exceptional accuracy with Stokes error below 0.01, and even lower at higher photon fluxes. We demonstrate this technology through diverse experimental scenarios, such as resolving structural details in biological tissues with a spatial resolution of 6~$\mu$m, characterizing rapid polarization transitions, and monitoring micro-scale birefringence in living or moving specimens.
	
\end{abstract}


\maketitle

\section{Introduction}

From imaging \cite{Jameson2010,Rimoli2022} and sensing \cite{Zhan2021} to ellipsometry \cite{Liu2021,Wen2024} and optical communications \cite{Wang2016,Chen2016}, light polarization plays a fundamental role across numerous scientific disciplines. In particular, the polarization of light emitted by an optically anisotropic specimen carries essential information about its material structure and optical properties. This capability has led to widespread applications in medicine, where polarization provides insight into the structural and morphological characteristics of various biological tissues and enables label-free detection of pathological changes \cite{Qi2017}. Polarization-resolved endoscopy and imaging techniques have proven valuable for detecting, classifying, and staging cancerous lesions in organs such as the stomach, colon, prostate, and larynx \cite{Doradla2013, Qi2023}. However, while polarization measurements can be relatively straightforward in a controlled laboratory environment, achieving highly accurate and reliable characterization under challenging conditions and in complex dynamic scenarios remains a significant obstacle.

\begin{figure}[ht]
	\centering
	\includegraphics[width=0.9\columnwidth]{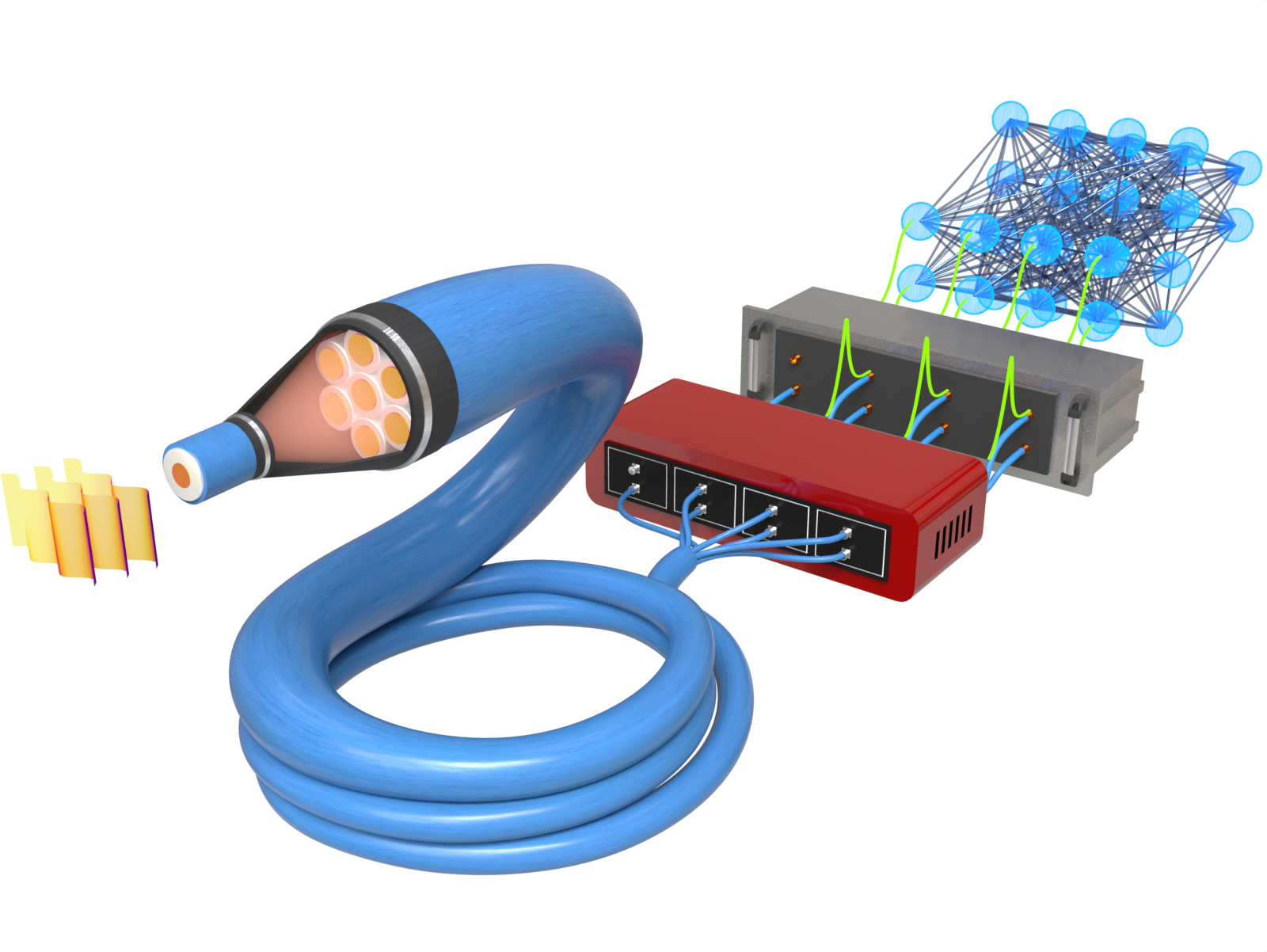}
	\caption{
		A visual representation of the all-fiber polarization sensor. Polarized light emitted from a specimen is collected by a short piece of a few-mode fiber. The intermodal interference generates a granular speckle pattern at the output. Several isolated samples of this pattern impinge a fiber array and propagate to a corresponding number of single-photon detectors. The recorded detections are electronically processed using deep learning methods to characterize the incident polarization state.
	}
	\label{Header}
\end{figure}

Low light intensities, representing the first challenging condition under consideration, come with a reduced signal-to-noise ratio, which complicates extracting meaningful polarization information from data burdened with environmental and shot noise. Similarly, conducting high-speed measurements, as the second condition, is accompanied by low signal levels within the acquisition time. While increasing the incident light intensity can help, it is limited by our ability to deliver the optical power to the particular sensing area and, most ultimately, by the photodamage threshold of the sample. Incorporating single-photon detectors can substantially improve the accuracy of protocols relying on the number of detected photons for polarization state reconstruction. However, another drawback arises as these detectors are relatively large and cannot be efficiently integrated. This issue becomes particularly problematic when faced with the third condition - constrained space. Such spaces not only impose restrictions on the overall size of the measurement setup but also limit the placement and alignment of optical components. Therefore, these challenging conditions pose a severe setback, especially for measurements during invasive medical procedures, such as microendoscopy \cite{Turtaev2018,Wen2023}, environmental monitoring \cite{Valentino2022,Behal2022}, or internal/in-situ inspections of materials \cite{Deng2021,Coppola2024}. We address these problems by developing a polarization sensor based on light propagation in disordered media.

Coherent light propagating within a disordered medium is subjected to multimode interference, generating a granular speckle pattern. Despite the apparent degrading effect, the pattern retains encoded information about the incident light~\cite{Rotter2017, Bertolotti2022, Cao2022, Vynck2023, Cao2023}. Analyzing the light propagation in disordered media and employing suitable post-processing techniques, one can study the disordering effects of the media~\cite{Valent2018, Matths2021}, estimate and classify input images~\cite{Cizmar2012, Plschner2015, Borhani2018, Rahmani2018, Caramazza2019, Wen2023}, learn to modulate the input for projecting a targeted image on the output~\cite{Rahmani2020}, reconstruct shapes of propagated ultrashort laser pulses~\cite{Ziv2020, Xiong2020}, compute and communicate~\cite{Defienne2016, Leedumrongwatthanakun2019, Valencia2020, Pierangeli2021, Courme2023}, and even make an accurate spectrometer or wavemeter~\cite{Redding2013, Gupta2020}. Recently, the speckle pattern has proved to preserve a great deal of information about the polarization state of the incident light~\cite{Juhl2019, Facchin2020}. Although additional studies successfully advancing the application of disordered-media propagation have been conducted~\cite{Rubin2022, Pierangeli2023, Stibrek2023, Goel2023}, developing a compact polarimeter capable of rapid measurements with single-photon sensitivity continues to be an open problem.

Here, we introduce a highly accurate, single-shot, all-fiber polarization sensor based on intermodal interference in a very short segment of a few-mode fiber. Instead of processing the complete spatial speckle pattern, our approach relies on its sparse sampling in a few isolated points using a fiber array. Combined with fiber-coupled single-photon detectors, this design allows for high sensitivity down to a single-photon level. The all-fiber sensor contains no moving components or complex metastructures and is sufficiently compact for procedures in constrained spaces while operating at a spatial resolution of 6~$\mu$m. Additionally, it supports a high-throughput operation capable of resolving several thousand polarization states per second. The setup is aided by a deep machine-learning model reconstructing complete polarization information in real-time. Once calibrated, the sensor maintains unparalleled accuracy, reaching Stokes errors well below 0.01, and long-term stability for over a month, even when measuring partially polarized light. We demonstrate its polarimetric capabilities through various experiments, including scans of dense connective tissue, birefringence measurements with a USAF test target, and dynamic polarization sensing in a moving diatom and during a fast twisted nematic liquid crystal transition. This all-fiber sensor sets new standards for compact, stable, and high-speed polarization measurements, offering a powerful tool for cutting-edge applications in biomedicine, material research, and beyond.


\section{Results}

\subsection{Experimental and data processing setup}\label{Sec_2}

The speckle-based polarization sensor illustrated in Fig.~\ref{Header} employs a common few-mode fiber to characterize the input polarization state. The sensor collects light with a front face of a 5~cm long step-index SMF28 fiber with a numerical aperture of 0.14 and a core diameter of 8.2~$\mu$m. While the motivation behind choosing this specific fiber type is to minimize the sensing area and keep the speckle structures large, we emphasize that this preference is optional. Virtually any few-mode or multimode fiber, including significantly shorter variants (see Sec.~\ref{Sec_2_2}), can be used as a substitute. The main requirement for the fiber is to support several interfering modes necessary for creating a polarization-dependent speckle pattern at the output. Following a 5mm free-space propagation, we collect discrete samples of this interference pattern using a fiber array comprising seven gradient multimode fibers, each with a 62.5~$\mu$m core diameter, forming a 375~$\mu$m diameter honeycomb pattern. Each fiber then propagates the collected intensity signal to an independent single-photon avalanche diode, recording the number of detection events. After normalizing the measured counts, this setup associates the sampled polarization state with the corresponding relative frequencies of each detector, which we term a count distribution. For more experimental details, see Sec.~\ref{Sub_SS1} and Fig.~\ref{Experimental_setup}.

Using the sensor, we analyzed 30,000 polarization states, uniformly covering the entire surface of the Poincar\'{e} sphere~\cite{huard1997polarization}. Our state preparation involved a continuous laser beam with a central wavelength of 0.8~$\mu$m attenuated to the single-photon level and propagated through a linear polarizer. Subsequently, a twisted nematic liquid crystal device~\cite{Bielak2021}, controlled by voltage signals, performed a targeted unitary transformation of the polarization state on this weak coherent light. The device operation underwent calibration through bidirectional modeling~\cite{Vainka2022}, ensuring fast and precise preparation of arbitrary polarization states. Its accuracy was verified using a reference polarimeter based on wave plates. The polarization state preparation using both methods reaches an average fidelity exceeding 0.999, however, the liquid crystal device operates by orders of magnitude faster. Subsequently, our sensor characterized the prepared polarization state, providing the corresponding detection counts. On average, we acquired approximately 150.000 photon detection events per detector within a 50~ms acquisition time window. Moreover, we expand our dataset to include partially polarized light. We establish a count distribution for a mixture of two orthogonal states by weight-summing their respective distributions. This approach is virtually equivalent to directly measuring the corresponding mixed state. Employing this method, we assembled an extensive dataset of polarization states distributed uniformly throughout the entire volume of the Poincar\'{e} sphere.

We used the dataset to train a fully connected deep neural network for reconstructing the polarization state given the associated count distribution, see Sec.~\ref{Sub_SS3} for details. The optimized and fully-trained network was evaluated on a test set of previously unobserved polarization states, also acquired from the experimental setup. We achieved an unprecedented average infidelity of $8 \times 10^{-4}$ with a $[5 \times 10^{-5}, 2 \times 10^{-3}]$ confidence interval using the 10th and 90th percentile. These results closely approach the fidelity limit imposed by the polarization preparation itself~\cite{Bielak2021}, indicating that our method reaches the fundamental accuracy bounds of the experiment. The achieved infidelity translates to a reconstruction error of approximately 0.01 per Stokes parameter. In contrast, previous methods based on random scattering typically reach errors around 0.05 while requiring bulky setups unsuitable for in-situ applications, achieving lower operational speed, lacking the single-photon sensitivity, and not presenting the ability to process partially polarized light~\cite{Juhl2019, Facchin2020, Rubin2022, Pierangeli2023}. To further validate our approach, we performed an additional benchmark using a 1951 USAF birefringent test. As detailed in Sec.~\ref{Sub_S2}, we compared our sensor with a polarization measurement based on rotating wave plates. The two methods yielded an average fidelity of approximately 0.987, underscoring the accuracy of our all-fiber sensor. Additionally, the USAF test also allowed the sensor spatial resolution estimation, which was determined to be approximately $6(5)~\mu$m. Together, these results comprehensively demonstrate the exceptional accuracy and robustness of our approach.


\subsection{Performance evaluation}\label{Sec_2_2}

\begin{figure}[h]
	\centering
	\includegraphics[width=0.99\columnwidth]{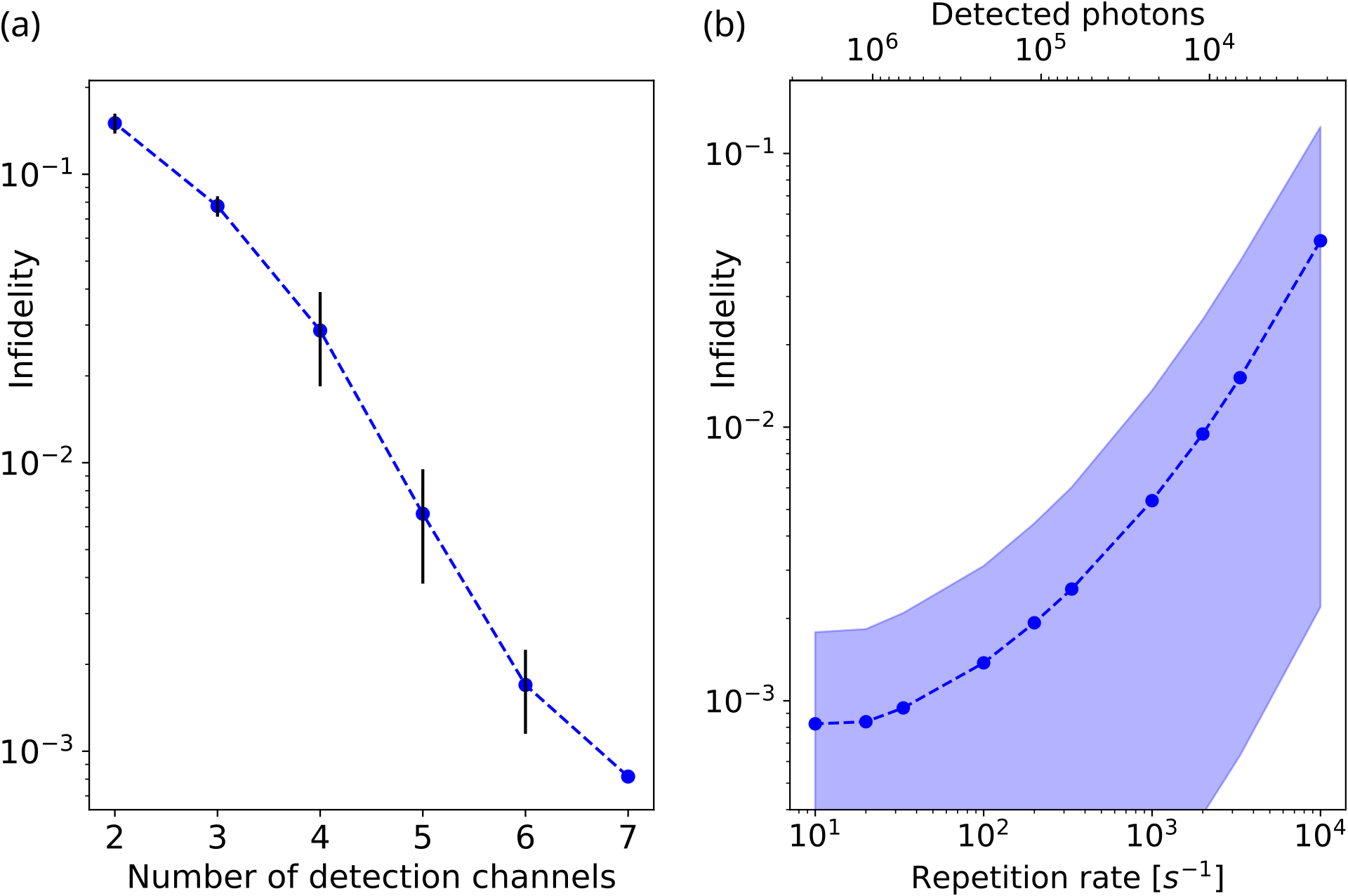}
	\caption{
		(a) Polarization error quantified by infidelity for varying numbers of active detectors. Variability bars represent uncertainties arising from different subset combinations. The achieved infidelities underscore the sufficiency of sparse sampling of the speckle pattern. (b) The relation between polarization infidelity, the collective number of detected photons across all detectors, and the measurement repetition rate. The colored area indicates the confidence interval of the experimental test set.
	}
	\label{Analysis}
\end{figure}

We highlight that our approach eliminates the need for capturing an image of the complete speckle pattern. Instead, we collect a limited number of speckle spots, which are then propagated through the fiber array channels. This substantial reduction in data collection allows for sensitive measurements at extremely low intensities using single-photon detectors. To assess the sensitivity of our approach to the processed portion of the speckle pattern, we investigate the achievable infidelity with regard to the number of employed detection channels. Following the same procedure, we trained additional networks utilizing only a subset of the detection channels as inputs. The infidelities achieved by these networks on corresponding test sets are depicted in Fig.~\ref{Analysis}~(a). The values represent the averaged errors across multiple trained networks, each employing a distinct combination of detection channel subsets. The results illustrated in this graph unequivocally demonstrate that extracting only a few samples from the speckle pattern provides sufficient information about the polarization state.

We also explored the dependence of the polarization sensing accuracy on the measurement repetition rate, i.e., the inverse value of the acquisition time, representing the number of polarization states characterized per second. Fig.~\ref{Analysis}~(b) displays the polarization infidelities of a single network, using all seven channels, evaluated for numerous repetition rates. This analysis allows us to determine the minimum required light intensity (i.e., the number of detected photons) needed to achieve a targeted infidelity value. Similarly, it allows estimation of the attainable fidelity for a given repetition rate. For example, the system achieves an average fidelity of 0.999 at a rate of 33~Hz - approximately the threshold at which the human eye perceives continuous motion. This result highlights the capability of our sensor for accurate, real-time polarization measurements. Furthermore, our setup demonstrates the capacity to measure over 2,000 polarization states per second using only $10^4$ photons while maintaining an average fidelity above 0.99. It is crucial to emphasize that these performances depend on the number of detected photons rather than the repetition rate alone. Consequently, even significantly higher repetition rates can attain the same fidelities with appropriately increased light intensity, limited virtually only by saturation of employed single-photon detectors. The presented photon efficiency makes the sensor particularly suitable for high-throughput, rapid-sensing applications.

\begin{figure}[h]
	\centering
	\includegraphics[width=0.99\columnwidth]{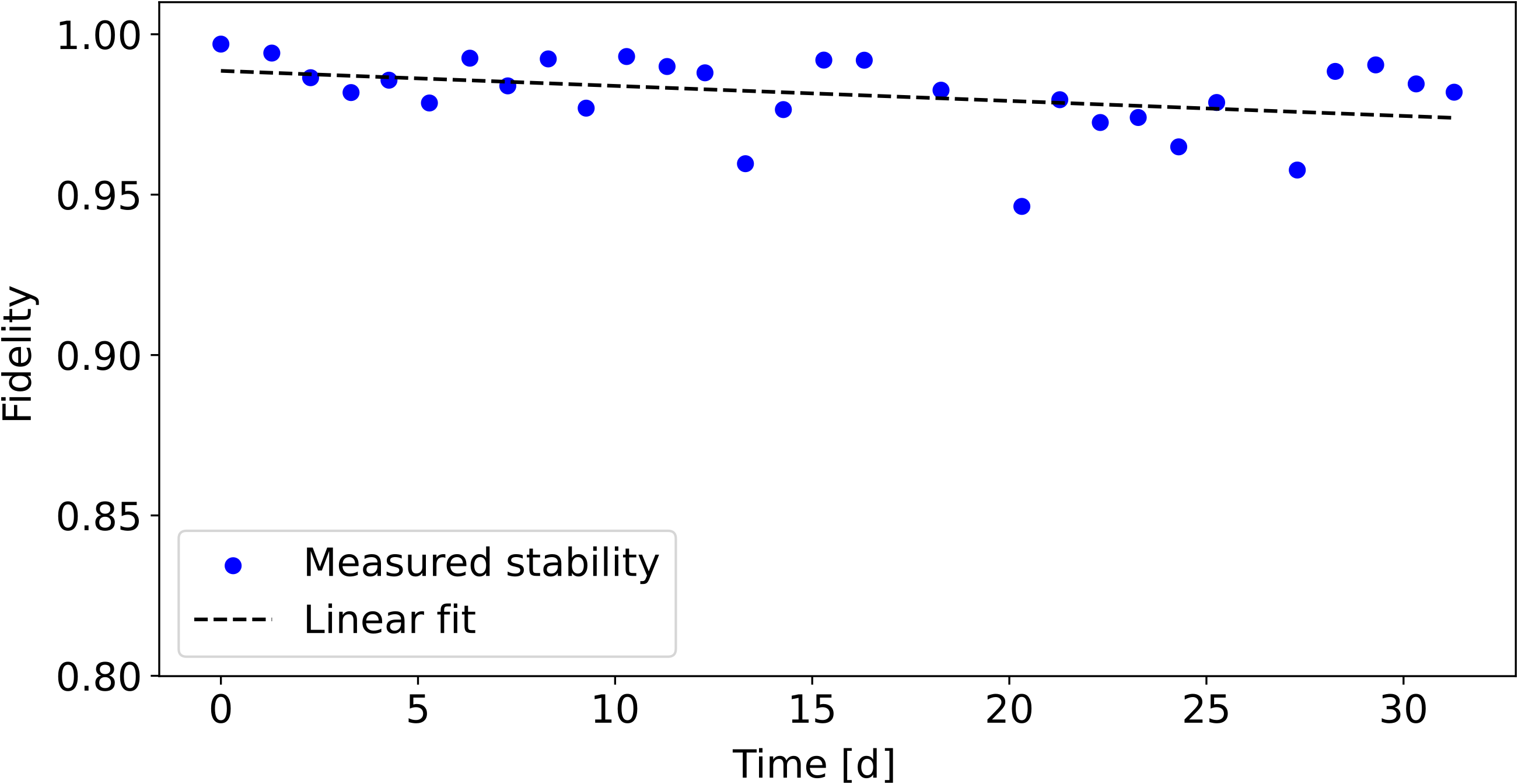}
	\caption{
		Long-term stability evaluation of the all-fiber polarization sensor employing a 12~mm long few-mode fiber encased in a ceramic ferrule. The system performance was evaluated periodically with a 24-hour repetition interval. The average fidelities (blue dots) are accompanied by a two-parameter linear fit (black line) with a $4.7$ $\times$ $10^{-4}~\text{day}^{-1}$ negative slope, capturing the slow, gradual decline.
	}
	\label{Stability}
\end{figure}

Furthermore, let us discuss the long-term stability of our approach and its robustness against environmental effects. Even when using a 5~cm long, few-mode fiber in an open environment without elaborated mechanical stabilization, the polarization sensor maintained sufficient stability for a week without recalibration or network retraining. We explored the possibility of further enhancing the stability using a 12~mm long fiber encased within a ceramic ferrule. This encasing rigidly fixes the fiber position relative to the fiber array, minimizing mechanical bending and reducing susceptibility to temperature gradients. With this modification, the system maintained its performance for over a month without recalibration. Fig.~\ref{Stability} illustrates the results of our long-term stability test, which was evaluated periodically at 24-hour intervals. The average fidelity was observed to degrade by approximately $4.7 \times 10^{-4}$ per day, i.e., $2 \times 10^{-5}$ per hour. This exceptional stability is attributed to three key factors: the short fiber length, which minimizes propagation-induced instabilities; the ferrule encasing, which mitigates environmental perturbations; and the use of a few-mode fiber, which produces a coarse-grained speckle pattern that increases the collection efficiency and reduces sensitivity to vibrations and drift. This design starkly contrasts with imaging-based applications, where long and massively multimode fibers have to be used with a deteriorating impact on the stability \cite{Cizmar2012,Plschner2015,Borhani2018,Rahmani2018,Caramazza2019,Wen2023}. Since sufficient intermodal coupling to generate a speckle pattern in a multimode fiber occurs over a propagation distance of several hundred wavelengths \cite{Wang2008, Wu2021}, we could reduce the employed fiber length even to sub-millimeter lengths. We expect that reducing the fiber to this scale could yield an additional order-of-magnitude improvement in stability. Finally, the sensor can be recalibrated to prolong its stability indefinitely. The recalibration process is straightforward, fully automatable, and can be completed in under an hour, making regular, even daily, recalibration a practical option. In summary, our polarization sensor offers a highly stable, accurate, and efficient solution for real-time, high-speed polarimetry with minimal maintenance requirements and strong resistance to environmental disturbances.

\subsection{Applications}

\begin{figure}[h]
	\centering
	\includegraphics[width=0.99\columnwidth]{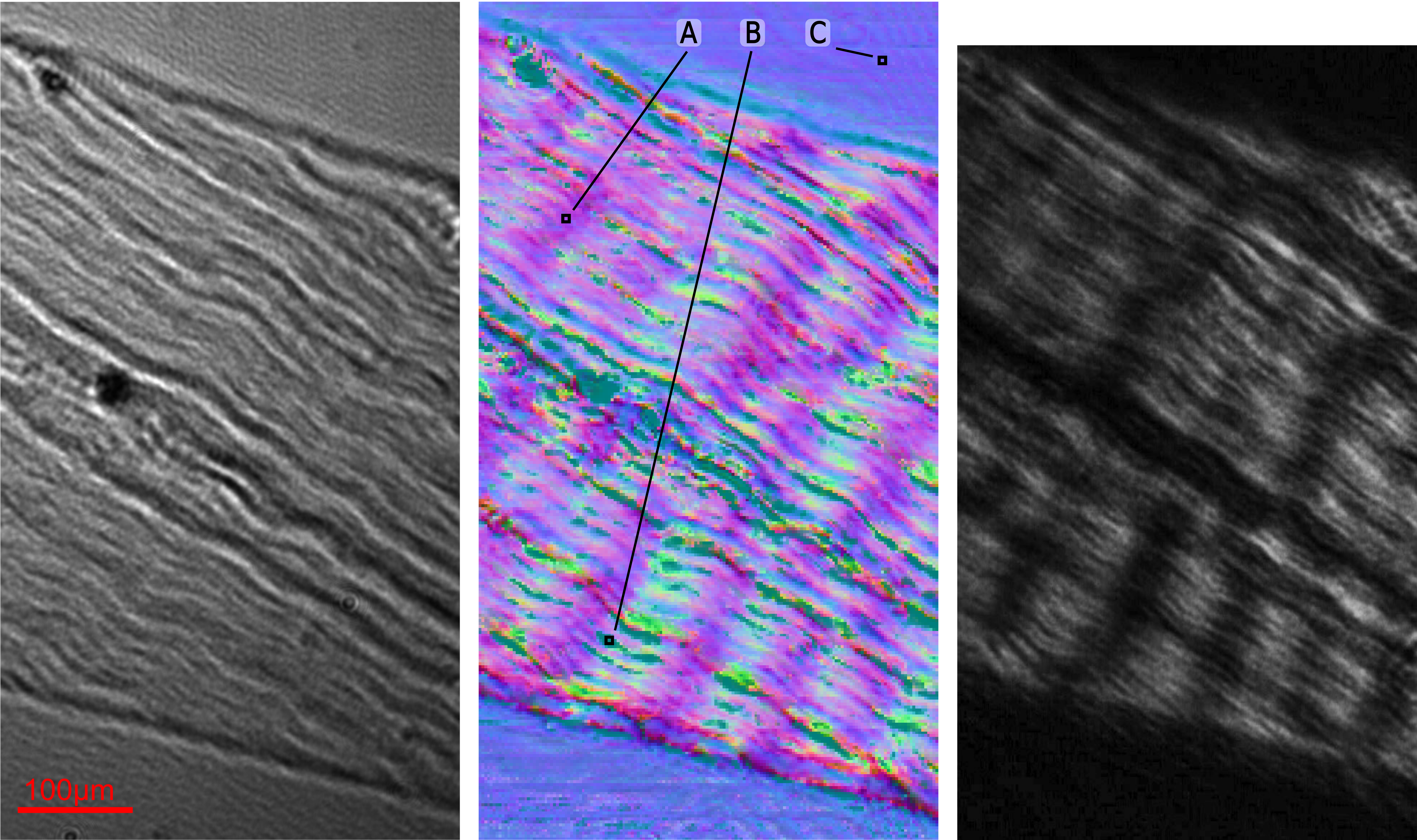}
	\caption{
		Visualization of dense connective tissue: (left) an intensity image, (middle) a scan using the all-fiber polarization sensor, (right) an image using a stand-alone polarization microscope. The resulting Bloch parameters of the all-fiber scan are represented as an RGB false-colored image. The three highlighted pixels characterize the purple and green segments in the polarization structure alongside a reference background polarization. Their respective Bloch parameters are 
		$\text{A}=(0.34, -0.43, 0.82),$ 
		$\text{B}=(-0.90, 0.05, 0.06),$ and 
		$\text{C}=(-0.04, -0.18, 0.98).$
	}
	\label{Sample_Imaging}
\end{figure}

We first demonstrate the sensing ability of our approach by conducting a polarization-sensitive scan of dense connective tissue (AmScope PS25W). While the primary application of our sensor significantly differs from polarization imaging, this optically anisotropic specimen is famously known in the polarization microscopy field. Visualizing its typical structures further underscores the accuracy of the presented polarimetric sensor. Using the experimental arrangement outlined in Sec.~\ref{Sub_SS1}, we replaced the calibration polarization preparation stage with the tissue specimen. The few-mode fiber was positioned proximal to the tissue, capturing the induced polarization transformation. We used motorized translation stages (Newport MFA-CC) with a bidirectional repeatability of $\pm 0.15~\mu$m and a maximum speed of $2500~\mu$m/s to scan over the tissue area. The reconstructed polarization states were expressed as three Bloch parameters, i.e., renormalized Stokes parameters~\cite{huard1997polarization}, and visualized as a false-colored RGB image, see Sec.~\ref{Sub_S2} for details. In Fig.~\ref{Sample_Imaging}, we present a side-by-side comparison of this polarization scan (middle) with an intensity image (left) captured from an identical tissue region. This intensity image was obtained using a monochrome camera (ImagingSource DMK~23U274) with a pixel size of $4.4 \times 4.4~\mu$m. We illuminated the specimen with 0.8~$\mu$m light and projected the resulting image onto the camera through a 25.4~mm focal-length lens, providing an estimated fivefold magnification. Moreover, the third panel (right) depicts a polarization image of the same specimen obtained using a stand-alone polarization microscope. Its experimental setup comprises a pair of crossed polarizers and the monochrome camera. As illustrated in Fig.~\ref{Sample_Imaging}, both polarization-based methods reveal underlying polarization structures that remain concealed in the intensity profile. This typical inner structure of the dense connective tissue verifies that our sensor performs polarization characterization, providing complete information about the polarization state.

\begin{figure}[h!]
	\centering
	\includegraphics[width=0.7\columnwidth]{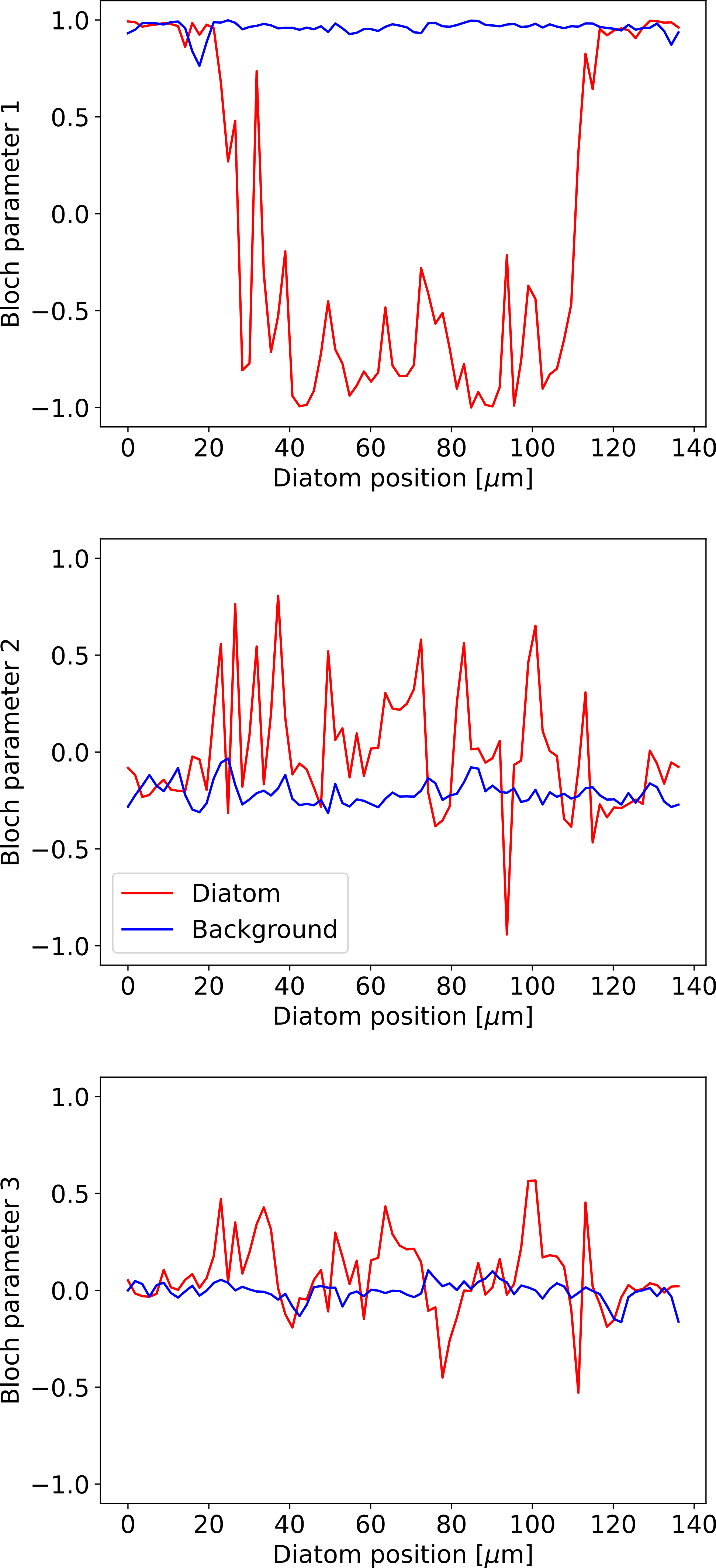}
	\caption{
		Spatial distribution of a diatom birefringent structure measured by the polarimetric all-fiber sensor and characterized using three Bloch parameters. The red color corresponds to changes induced by an anisotropic diatom passing in front of the fiber tip. For comparison, the blue line represents the same environment without the diatom. The modulation visible in all three Bloch parameters indicates the properties of a polarization-affecting element moving in front of the sensor, allowing further study of the specimen.
	}
	\label{Moving_Diatom}
\end{figure}

As the dense connective tissue is a static specimen, the complete polarization analysis provided by the all-fiber sensor can also be performed using an advanced imaging polarimeter comprising rotating wave plates. However, such a technique is insufficient for characterizing a rapidly changing polarization state induced by moving or evolving specimens. These scenarios require significantly faster approaches that have to be performed in a single-shot regime. To demonstrate the dynamic capabilities of our all-fiber sensor, we conducted polarization measurements of two systems: a floating diatom and twisted nematic liquid crystals undergoing a fast voltage-controlled transition. Starting with diatoms, these microscopic unicellular organisms comprise a complex inner structure that can exhibit anisotropic properties. We characterized the birefringence of an Actinoptychus heliopelta diatom positioned within the same experimental arrangement as the previous application. The spatial variation in the Bloch parameters across the diatom is depicted in Fig.~\ref{Moving_Diatom} with red lines. Compared to the blue lines of background polarization, the diatom properties are visible in the modulation of Bloch parameters. This analysis can count the number of living specimens, estimate their size and morphology, and classify them based on their birefringent properties. Beyond biological samples, the sensor can also find applications for fast in-situ material inspection, local strain analysis, and micro crystal growth monitoring.

\begin{figure}[h]
	\centering
	\includegraphics[width=0.7\columnwidth]{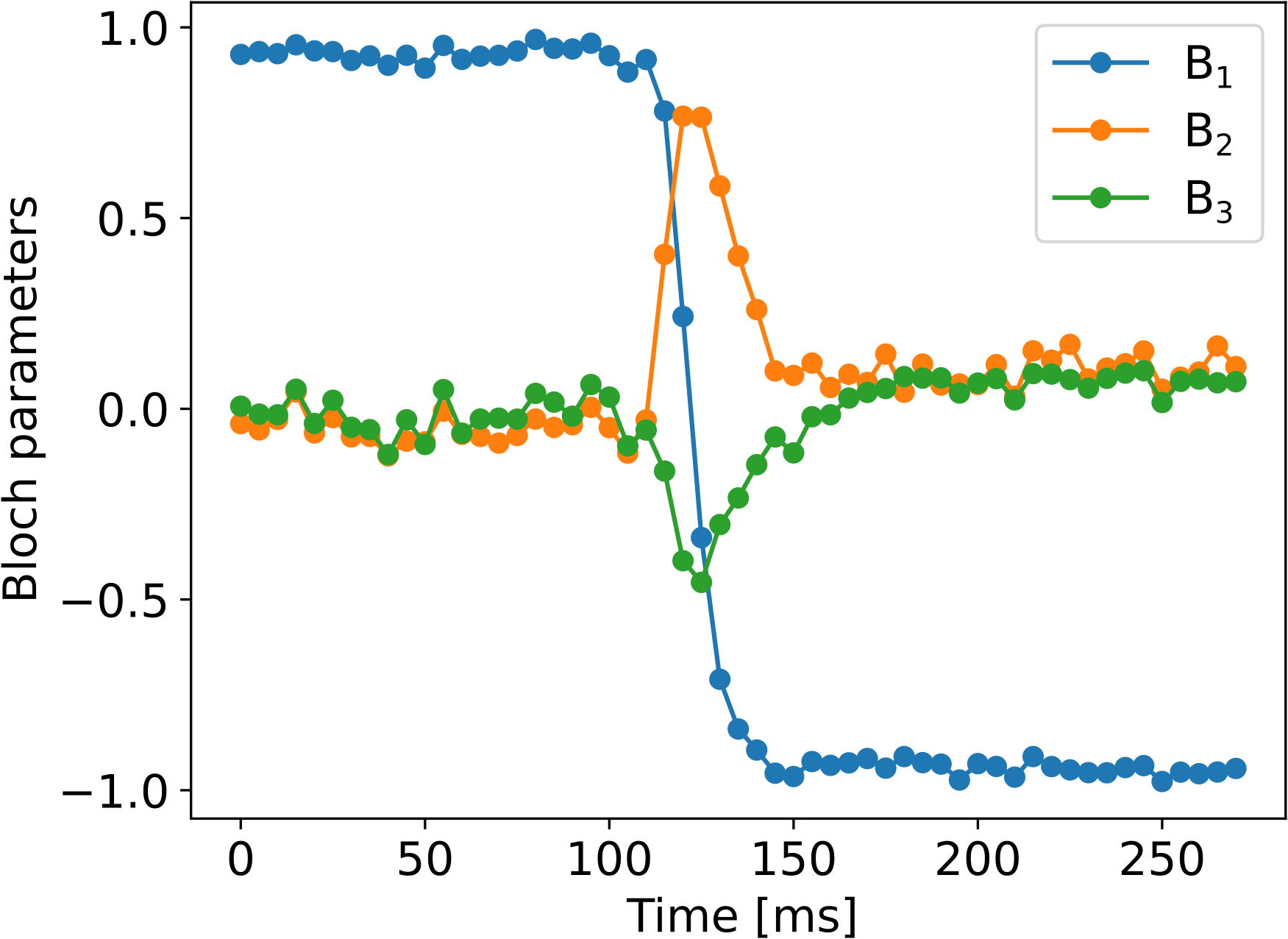}
	\caption{
		Time-resolved measurement of a polarization transition induced by voltage-controlled twisted nematic liquid crystals. The all-fiber sensor continuously captures the complete polarization information with a 5~ms acquisition window at the approximate power level of $10^5$ detected photons. The transition from horizontal to vertical polarization is represented through the evolution of the three Bloch parameters, each shown in a different color.
	}
	\label{TNLC_transition}
\end{figure}

Finally, we demonstrate the high-speed polarimetric capabilities of our sensor using a rapidly evolving specimen: twisted nematic liquid crystals. They allow for fast polarization state transformation, controlled by low-voltage signals that manipulate their spatial alignment. During reorientation, the induced polarization transformation undergoes a rapid transition, typically within a time window ranging from 50 to 600~ms \cite{Bielak2021}. We characterized the horizontal-to-vertical transition by continuously measuring the prepared polarization state using a 5~ms acquisition window. The reconstructed Bloch parameters corresponding to this process are visualized in Fig.~\ref{TNLC_transition}. The first Block parameter exhibits a smooth evolution from $B_1=1$ (horizontal polarization) to $B_1=-1$ (vertical polarization). The remaining two parameters reveal that the transition corresponds to a continuous rotation on a Bloch sphere, passing close to the diagonal state ($B_2=1$). The complete transition occurs within approximately 50~ms, validating the ability of our sensor to accurately capture dynamic polarization changes in real time, which is inaccessible to common rotating-wave-plate polarimeters. This demonstration, alongside the previously discussed applications, underscores the versatility and performance of our all-fiber polarization sensor for fast, low-light, high-precision polarimetry.

\section{Discussion}

We have developed a single-shot polarization sensor based on light propagation through a short rigid piece of a few-mode fiber, resulting in polarization-dependent scattering. This process provides a coarse speckle pattern with low signal losses and unprecedented long-term stability, which is virtually unaffected by fiber bending. The defining aspect of our approach is the sparse sampling of this pattern using a fiber array instead of capturing the whole intensity profile. This characteristic allows the use of any photonic detectors, including high-speed or single-photon detectors, enabling fast response and operation at low light intensities down to the picowatt level. The implementation of deep learning algorithms in the system ensures an accurate reconstruction, providing complete information about the polarization state, including partial polarization, with an infidelity of $8 \times 10^{-4}$. This superior performance remains stable for over a month without requiring adjustments or recalibration. Complemented by the rapid operation speed of more than two thousand high-fidelity reconstructions per second, it allows for both real-time and high-repetition-rate polarization sensing with remarkable precision. The all-fiber sensor, featuring a compact design with no moving components, is particularly well-suited for reducing the invasiveness of biomedical procedures in constrained spaces and under low-light conditions. Furthermore, one can enhance the spatial resolution of our sensor by tapering the tip of the few-mode fiber to reach the limit on near-field sensing while preserving the single-photon polarization sensitivity. However, there will be a trade-off between the increased spatial resolution and decreased collection efficiency. Also, the stability performance might be negatively affected.

In the view of optically-assisted machine learning, the random propagation in multimode fiber followed by a computational neural network represents an instance of an extreme learning machine or reservoir computing \cite{Pierangeli2021, Mujal2021}. By demonstrating the single-photon sensing capability of these architectures, we have provided the foundation for decreasing energy consumption in optical sensing, similar to what has been studied recently for computing \cite{McMahon2022}. Furthermore, to fully quantify the flow of information about the measured quantity (polarization state of light in our case), one may contemplate utilizing the Fisher information concept that has been only recently adapted for wave scattering \cite{Rotter2024}. The crucially missing part, and possible future focus of the research community, might be the analysis of the flow of Fisher information in neural networks \cite{Rotter2023Obergurgel,Rotter2024SPIE}. This would complement the analysis of the physical part of the sensor and allow for the full theoretical information analysis of the deep-learning aided metrology based on random wave scattering.

In summary, we presented a novel polarization sensor that is explicitly demonstrated for arbitrary polarization states, including partially polarized light, and simultaneously offers: single-photon sensitivity (picowatt powers transmitting a sample); rapid operation with thousands of reconstructions per second; high accuracy reaching Stokes errors below 0.01, and even lower at higher photon fluxes; long-term stability for over a month, unprecedented among imagers and sensors using multimode fibers; high spatial resolution of 6~$\mu$m; and a compact, all-fiber design without moving components. In addition, we analyzed the resource budget of the sensor and conducted a thorough confidence analysis, which is often neglected for deep learning applications.
Our polarization sensor significantly advances the field, answering the demand for a combination of compactness, accuracy, low light sensitivity, and high operational speed, which opens up new possibilities for highly accurate polarization sensing across various applications. 
Furthermore, our contribution lays down novel research pathways to single-photon-level sensing using random optical networks and media. It also poses new questions on information and confidence evaluation of systems consisting of random media together with deep neural networks. The ability to extract meaningful data from minimal input has far-reaching implications in fields where resource efficiency and sensitivity are paramount.


\section{Materials and methods}\label{sec11}

\subsection{Experimental setup}\label{Sub_SS1}

\begin{figure}[h]
	\centering
	\includegraphics[width=0.99\linewidth]{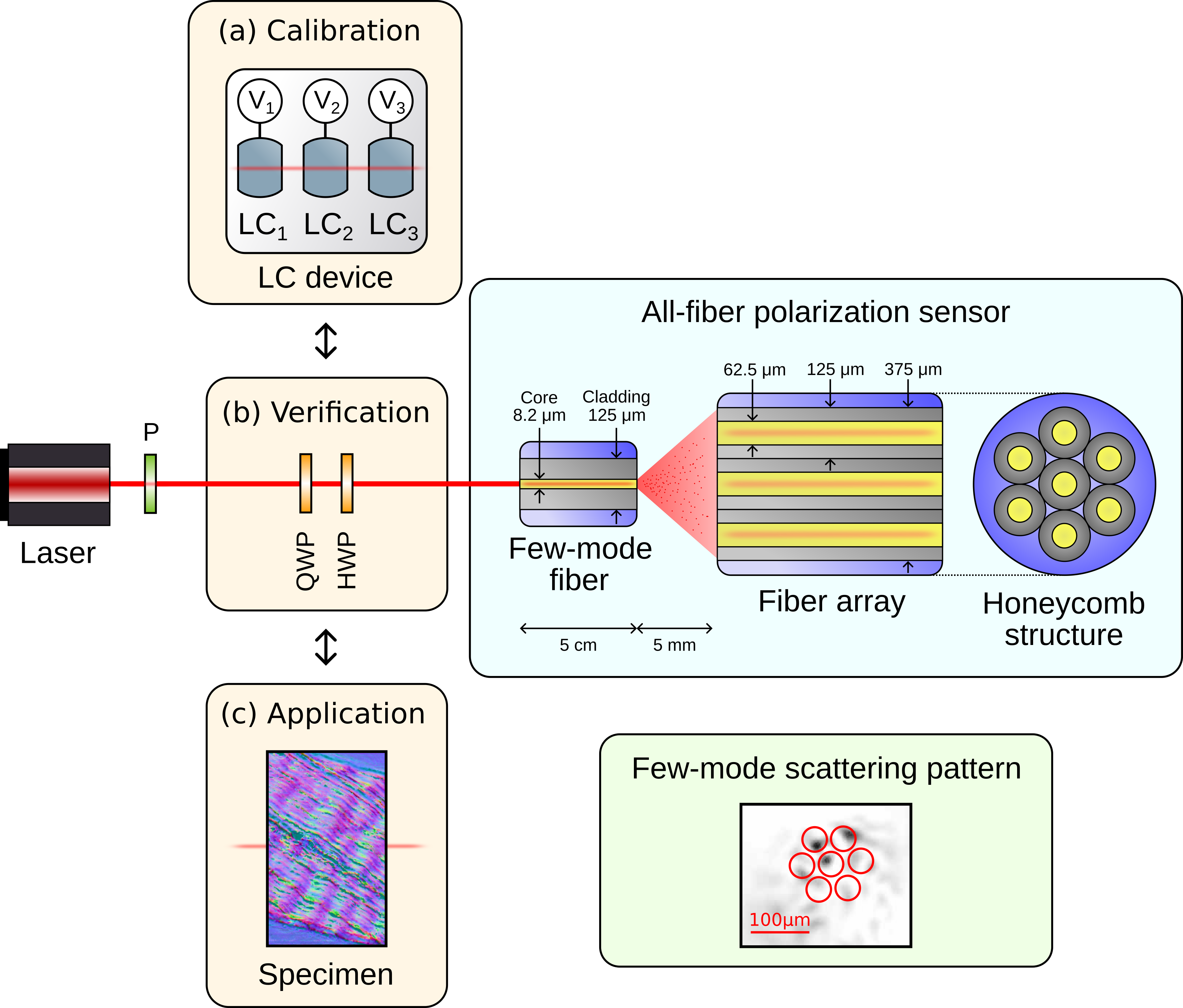}
	\caption{A detailed visualization of the experimental setup for the all-fiber sensor. The twisted nematic liquid crystal device, labeled as (a), is present only during dataset acquisition for deep neural network training. In the verification stage, depicted in (b), a reference polarimeter based on wave plates is employed to validate the accurate operation of the liquid crystal device. Finally, this segment is replaced with a polarization-modulating specimen (c) during the USAF test, dense connective tissue, diatom measurements, and fast liquid crystal transition. The green panel depicts a polarization-dependent speckle pattern with inverted intensities generated by the few-mode fiber. The red circles represent the sparse areas sampled by the subsequent fiber array.}
	\label{Experimental_setup}
\end{figure}

In addition to the information provided in the main text, we include a detailed scheme of our all-fiber polarization sensor, depicted in Fig.~\ref{Experimental_setup}. This scheme comprises three configurations. In the first configuration, the twisted nematic liquid crystal device~\cite{Bielak2021} induces polarization transformation on the attenuated coherent signal generated by a fiber-coupled continuous 810~nm semiconductor laser diode (QPhotonics QFLD). The laser diode exhibits typical fluctuations, drifts, and mode hops within a 1~nm wavelength range. With a more stable source, such as a distributed-feedback laser diode or solid-state laser, the wavelength stability would increase significantly, which would completely neglect any negative effects on the performance of our sensor. Moreover, the sensor can be recalibrated for an arbitrary wavelength within an hour. We independently verified the accurate operation of the liquid crystal device in the second configuration by replacing it with a reference polarimeter based on rotating wave plates. The third configuration represents the sensor application and comprises the birefringent specimen under investigation. In all cases, the modified signal is collected by the 5~cm long SMF28 few-mode fiber. The number of modes supported by this fiber is characterized in Sec.~\ref{Sub_SS2}.

The generated speckle pattern propagates through a 5~mm free space before reaching the array of seven gradient multimode fibers GIF625 (SQS Vláknová optika, 62.5/125/250$\mu$m MM OM1 Fiber). This propagation allows the speckle grain size to match the multimode fiber core. Arranged in a honeycomb structure with a 375~$\mu$m diameter, the multimode fibers have a core diameter of 62.5~$\mu$m and a cladding diameter of 125~$\mu$m. Each fiber sparsely samples the intensity from the speckle pattern and forwards the signal to an independent single-photon avalanche diode (Excelitas SPCM). The entire experimental setup is located in an unsealed open space of a temperature-maintained laboratory, with the temperature varying approximately within 1.5~°C and the relative humidity ranging from 25\% to 35\% during the measurements. Despite these fluctuations, we demonstrated unprecedented stability and accuracy over many weeks (see Section \ref{Sec_2_2}). This level of robustness is fully sufficient for any laboratory measurements. Moreover, in typical biomedical and material research, the environmental conditions, such as sample temperature, are deliberately controlled and stabilized. Under such conditions, the long-term stability of our sensor will be even greater than demonstrated.

\subsection{Supported fiber modes analysis}\label{Sub_SS2}

Determining the exact number of modes $M$ supported by the SMF28 few-mode fiber at the given wavelength proves intricate. Most mathematical formulas are precise only in the limit case of a high mode number. Nevertheless, we estimated $M$ by evaluating the normalized frequency parameter~$V,$
\begin{equation}
	V = \frac{2 \pi}{\lambda} r_{\text{core}} \text{NA},
	\label{V_param}
\end{equation}
where $\lambda = 810~\text{nm}$ is the wavelength of propagated light, $r_{\text{core}} = 4.1~\mu \text{m}$ is the fiber core radius, and $\text{NA} = 0.14$ is the numerical aperture. The calculated value of $V = 4.45 > 2.405$ indicates that the fiber supports more than one mode per polarization direction. Subsequently, the number of modes $M$ is approximately given by the relation
\begin{equation}
	M \approx \frac{V^2}{2} = 9.9,
	\label{Num_modes}
\end{equation}
valid under the assumption that $V$ is large. Since this condition was not fully satisfied, we have also utilized publicly available software developed for mode computation in multimode fibers~\cite{Replicate} and obtained the number of modes $M = 8$.


\subsection{Numerical setup}\label{Sub_SS3}

The physical setup described above is assisted by a numerical processing comprising a deep neural network model. This trained network, consisting of 250 neurons per 4 hidden layers, performs a nonlinear transformation on the count distribution and returns four real-valued outputs. These represent elements of a two-by-two triangular matrix $\tau$ with real-valued diagonal entries and a single complex off-diagonal entry. Utilizing the Cholesky decomposition, we can reconstruct a Hermitian positive-definitive matrix $\tau \tau^\dagger.$ Upon normalizing its trace to unity, $\rho = \frac{\tau \tau^\dagger}{\text{Tr}\left[ \tau \tau^\dagger \right]}$ represents a physically sound coherence matrix of the polarization state~\cite{huard1997polarization}, mathematically equivalent to a density matrix of a two-level quantum system. Therefore, the complete model can be interpreted as providing the polarization coherence matrix of the collected light given the measured count distribution. Alternatively, the polarization state can be described by the Bloch parameters $\left( B_1, B_2, B_3 \right)$, $\rho = \frac{1}{2} \left( 1 + \sum_{j=1}^{3}{B_j \sigma_j} \right)$, where $\sigma_j$ are Pauli matrices.

We trained the network using the Adam optimizer and a mean squared error loss function. Various hyperparameters, including network architecture, were fine-tuned to optimize the network performance. For this purpose, we employed a validation set, separated from the training data, to find the optimal hyperparameter combination with respect to a fidelity metric 
\begin{equation}
	F = \text{Tr} \left[ \sqrt{\sqrt{\rho} \cdot \sigma \cdot \sqrt{\rho}} \right] ^2 \in \left[ 0,1 \right],
	\label{Fidelity}
\end{equation}
where $\rho$ is the polarization coherence matrix provided by the network and $\sigma$ is the ground truth matrix of the input polarization state. This metric, which quantifies the closeness between the states, can be expressed in terms of error as infidelity $1-F$. Finally, the separated experimentally acquired test set was used for the performance evaluation of the optimal network.


\subsection{1951 USAF birefringent test}\label{Sub_S2}

\begin{figure}[h]
	\centering
	\includegraphics[width=0.99\linewidth]{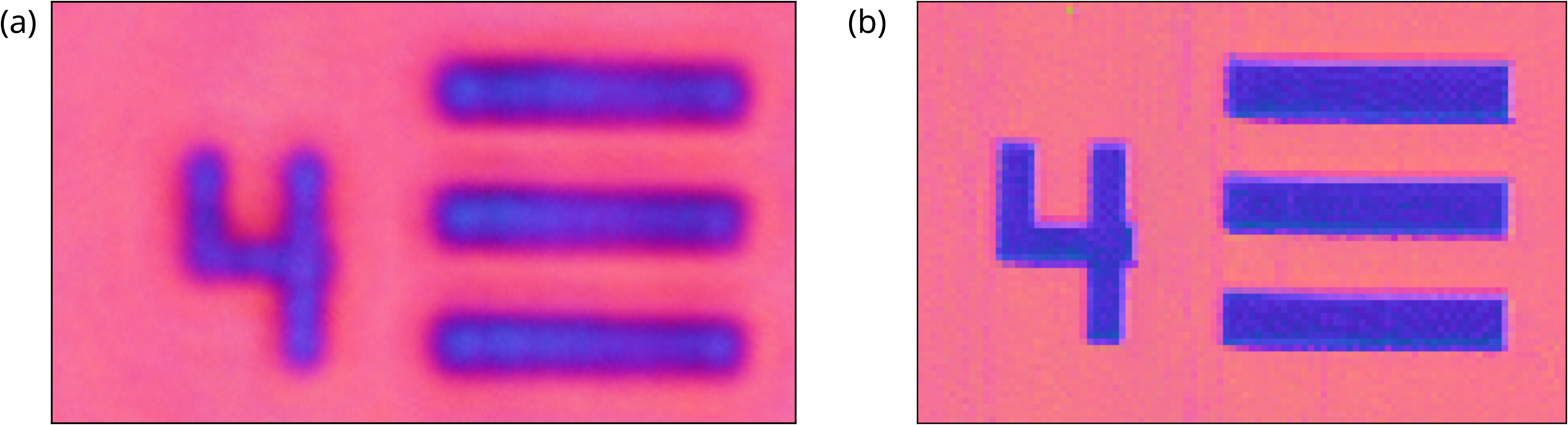}
	\caption{A comparison of polarization measurement on a region of the 1951 USAF birefringent resolution test. (a) A polarization image obtained using standard imaging based on rotating wave plates. (b) An all-fiber sensor scan of the same region. Both results are visualized as false-colored RGB images using Bloch parameters.}
	\label{USAF_test}
\end{figure}

We conducted a scan of a 1951 USAF birefringent resolution test with the developed polarization sensor and compared it with a polarization measurement based on rotating wave plates. Fig.~\ref{USAF_test} shows a polarization image obtained with the standard method (a) alongside the scan using our all-fiber sensor (b). Both panels visualize false-colored RGB images using Bloch parameters $\left( B_1, B_2, B_3 \right)$ with $R = \frac{B_1 + 1}{2}, G = \frac{B_2 + 1}{2},$ and $B = \frac{B_3 + 1}{2}$ corresponding to red, green, and blue channel intensities. By averaging over the background and polarization-modulating regions, we estimated a fidelity value of approximately 0.987 between the two polarization measurement approaches. This comparison unequivocally demonstrated the ability of our sensor to perform highly accurate polarization measurements.

Additionally, the image shown in panel (b) allows for the evaluation of the spatial resolution of our sensor. To this end, we fitted horizontally and vertically oriented edges at several positions. Characterized as the 20\% to 80\% width, we estimated the resolution to be approximately $6(5)~\mu$m. Besides the distance from a specimen to the fiber tip, the resolution is dictated by the few-mode fiber diameter and can be modified by employing different fiber types. Moreover, for applications requiring higher resolution, utilizing fiber tip tapering could improve it up to the near-field scanning optical microscopy at the cost of the collection efficiency and possibly stability.

\medskip
\textbf{Acknowledgements} \par

We acknowledge the use of cluster computing resources provided by the Department of Optics, Palack{\'y} University Olomouc. We thank Jan Provazn{\'i}k for maintaining the cluster and providing support. We also thank Josef Hlou{\v{s}}ek for developing a 3D model shown in Fig.~\ref{Header} and Jarom{\'i}r B{\v{e}}hal for a fruitful discussion on birefringent biological samples.

\medskip
\textbf{Availability of data and materials} \par

The code and data that support the findings of this study are publicly available on GitHub: \\\url{https://github.com/VasinkaD/Polarization-Deep-Sense}.

\medskip
\textbf{Funding} \par

This work was supported by the Czech Science Foundation (project 21-18545S), Ministry of Education, Youth, and Sports of the Czech Republic (project OP JAC CZ.02.01.01/00/23\textunderscore021/0008790).
MB and DV acknowledge the support by Palack{\'y} University (projects IGA-PrF-2024-008 and IGA-PrF-2025-010).

\medskip
\textbf{Authors' contributions} \par

MB assembled the sensor and conducted the experiments. DV wrote the deep learning and data processing codes, and drafted manuscript. MJ initiated and supervised the project. All authors contributed in data interpretation and revising manuscript.

\medskip
\textbf{Conflict of Interest} \par

The authors declare that they have no conflict of interest.

\bibliography{refs}

\end{document}